\documentclass[a4paper]{article}
\usepackage{physics}
\usepackage{fancyvrb}
\newcommand{\ee}[1]{\mathrm{e}^{#1}}
\newcommand{\delh}{\mathnormal{\Delta}H}
\newcommand{\pz}{P}
\newcommand{\pbar}{Q}
\usepackage{orcidlink}

\usepackage[]{cite}
\usepackage{amsmath,amssymb,amsfonts}
\usepackage[margin=1in]{geometry}
\RequirePackage{graphicx}
\usepackage{authblk}

\numberwithin{equation}{section}
\title{An explicit formula for perturbation theory at any order with infinitely many perturbations}

\author{Joseph M. Jones\thanks{jxj898@theory.bham.ac.uk}\, and M. W. Long}

\affil{School of Physics and Astronomy, University of Birmingham, Edgbaston, Birmingham, B15 2TT, United Kingdom}

\begin{document}
	\maketitle
	
	\begin{abstract}
		We provide a systematic formula, in terms of integer partitions, that generates perturbation theory explicitly at an arbitrary order.
		Our approach naturally includes an infinite number of perturbations and uses a single matrix equation that contains the information for both the eigenvalue and eigenvector corrections.
		The formula reduces to the standard case of one perturbation in the appropriate limit.
		This formulation streamlines the derivations that are traditionally tedious in perturbation theory, facilitating high-order calculations.
	\end{abstract}
	
		\section{Introduction}
		Perturbation theory is a cornerstone of theoretical physics, providing approximate solutions when exact results are inaccessible. Despite its ubiquity, it is usually only employed to a low---often second---order. Explicit formulae for corrections at arbitrary order are rarely written down, largely due to the algebraic complexity of higher-order terms. In this work, we introduce a systematic approach that resolves this difficulty by expressing perturbation theory in terms of integer partitions.
		
		Our method yields closed-form expressions for corrections at any order and naturally accommodates an infinite set of perturbations. The result is compact and algorithmically tractable, allowing both eigenvalue and eigenvector corrections to be obtained from a single matrix equation.
	
	Our approach is analogous to L\"owdin’s matrix formulation of perturbation theory \cite{lowdin1962}, but extends it by including infinite perturbations.
	Previous work on high-order perturbation theory has explored various aspects of the problem \cite{kato1949,messiah2014,brueckner1955,lowdin1962,soliverez1969,bracci2012}; we summarise these developments and their relation to our approach later in this introduction.
	
	As a physical example of where one may encounter an infinite number of perturbing Hamiltonians, one may consider the recently discovered power series representation of the Baker–Campbell–Hausdorff formula \cite{moodie2021,jones2025a},
	\begin{align*}
		\ee{2C} &= \ee{A}\ee{2B}\ee{A}\\
		\Rightarrow 
		C &= A + \sum_{N=1}^\infty G_N(L_1,L_2,\dots,L_N)\, B_1B_2\cdots B_N,
	\end{align*}
	for non-commuting $A$ and $B$, which expresses $C$ as a power series in $B$, with the $A$-dependence resummed exactly into the hyperbolic coefficients $G_N$ at each order. The commutator operators $L B\equiv[A, B]$ are indexed to label which $B$ they operate on in the ordered product, for example,
	\begin{align*}
		L_mB_1B_2\cdots B_{m-1}B_mB_{m+1}\cdots B_{N-1}B_N=\\
		B_1B_2\cdots B_{m-1}[A,B_m]B_{m+1}\cdots B_{N-1}B_N.
	\end{align*}
	For diagonal $A$ and non-diagonal $B$, $A$ may be viewed as the exactly solvable reference point for a perturbative expansion, with $G_NB^N$ representing the $N$'th order perturbation. The idea of developing perturbation theory for this infinite series provided the physical motivation for the present work and will be explored further in \cite{jones2025b}.
	
	The most common formulations of perturbation theory are Rayleigh-Schr\"odinger perturbation theory (RSPT) \cite{rayleigh1894,schrodinger1926} and the Brillouin-Wigner perturbation theory (BWPT) \cite{lennardjones1930, brillouin1932, wigner1997}.
	BWPT provides a self-consistent formula that must be solved numerically. 
	RSPT provides a formula that can be evaluated by knowing only the previous corrections to the energy.
	
	Our result is for RSPT but appears more like the derivation of BWPT; as we use a recursive formula but importantly we do not end up with a self-consistent equation. Our result is provided in terms of ordered integer partitions \footnote{We use integer partitions where the ordering matters, for example the ordered partitions of $4$ are $\{(4)$, $(3,1)$, $(1,3)$, $(2,2)$, $(2,1,1)$, $(1,2,1)$, $(1,1,2)$, $(1,1,1,1)\}$. These are sometimes referred to as compositions \cite{stanley2011} by the mathematical community.} and we provide the explicit formula for the corrections at a particular order. Both BWPT and degenerate RSPT could also be derived in this formalism as an extension to L\"owdin's work \cite{lowdin1962} but that is beyond the scope of this paper.
	We consider time-independent Hamiltonians with non-degenerate energy levels, and ignore the question of convergence as this is only relevant to applications of the theory to a physical model.
	
	We believe that there are two interesting results in this work:	
	\begin{enumerate}
		\item We formulate perturbation theory for an infinite number of perturbing Hamiltonians, generalising the standard single-perturbation case, which is recovered in the appropriate limit.
		The result is a streamlined derivation of perturbation theory and a ready-to-use formula that can be directly applied to physical problems.
		\item We derive an explicit expression for both the eigenvalue and eigenvector corrections at any order as a sum over integer partitions. Each partition specifies how perturbations from the infinite set are selected at a given order and simultaneously determines which lower-order energies contribute and where.
	\end{enumerate}
	
	\subsection{Main result}
	To state the main result we require some definitions and assumptions. Firstly, define the following generating functions
	\begin{align*}
		H(z)&\equiv H_0+\sum_{n=1}^{\infty}H^{(n)}z^n,\,\, 
		E(z)\equiv \epsilon_0+\sum_{n=1}^{\infty}E^{(n)} z^n,\\ 
		\psi(z)&\equiv \ket{0}+\sum_{n=1}^{\infty}\ket*{\psi^{(n)}}z^n,
	\end{align*}
	where $H^{(n)}$, $E^{(n)}$ and $\ket*{\psi^{(n)}}$ are the $n$'th order perturbation, eigenvalue correction, and eigenvector correction, respectively, and $z^n$ controls the order of the expansion in the later generating function analysis.
	
	Perturbation theory starts with an exactly solved problem and perturbs around this. For us the exactly solved problem is $H_0\ket{0}=\epsilon_{0}\ket{0}$ where all unperturbed states eigenvalues and eigenvectors are assumed to be known.
	We choose $\ket{0}$ as the unperturbed \textit{starting state} which has energy $\epsilon_0$, without loss of generality.
	We have the usual assumptions for doing perturbation theory; the unperturbed kets are mutually orthonormal, the perturbed kets are not necessarily normalised, and we choose for the eigenvector corrections to contain none of the starting state, $\pz\ket*{\psi^{(n)}}=0$, where $\pz\equiv\dyad{0}$, $\pz^2=\pz$, is a projector that projects onto the starting state.
	We also have its complement $\pbar\equiv 1-\pz$ that projects out the starting state, $\ket*{\psi^{(n)}}=\pbar\ket*{\psi^{(n)}}$. The central quantity in our analysis is
	\begin{align*}
		\delh^{(n)} &\equiv H^{(n)}-E^{(n)}\pbar,\quad n>0.
	\end{align*}
	Including both contributions within a single quantity leads to a more compact formulation \footnote{For readers familiar with high-order expansions, the physical motivation for defining $\Delta H^{(n)}$ can be interpreted using a disconnected-cluster argument \cite{dyson1949,kubo1962}. The subtraction of $E^{(n)}$ removes disconnected, non-extensive contributions to the final energy, ensuring that only connected components of the perturbation survive at each order.} than is typically found in the literature, even for a single perturbation \cite{soliverez1969,bracci2012}.
	
	Finally we require the denominator matrix
	\begin{align*}
		\Gamma\equiv \pbar(\epsilon_0-H_0)^{-1}=(\epsilon_0-H_0)^{-1}\pbar=\sum_{n_1\neq 0}\frac{\dyad{n_1}}{\epsilon_0-\epsilon_{n_1}}
	\end{align*}
	where the final equality is the spectral representation of $\Gamma$.
	Perturbation theory with an infinite set of perturbations then amounts to solving the following equation
	\begin{align}
		H(z)\psi(z)=E(z)\psi(z)\label{eq:def}
	\end{align}
	recursively for $E^{(N)}$. Indeed, one requires all previous orders to solve for the next eigenvalue and eigenvector corrections so it is necessarily an iterative process.
	We find that the corrections to the eigenvalue and eigenvector at $N$'th order are
	\begin{equation}\label{eq:res}
		\begin{aligned}
			E^{(N)} &= \sum_{p=1}^{N}\sum_{n_1=1}^{\infty}\cdots\sum_{n_p=1}^{\infty}\delta_{n_1+\cdots+n_p,N}\mel*{0}{H^{(n_1)}\Gamma \delh^{(n_2)}\cdots\Gamma \delh^{(n_{p-1})}\Gamma H^{(n_p)}}{0}\\
			\ket*{\psi^{(N)}}
			&=\Gamma \sum_{p=1}^{N}\sum_{n_1=1}^{\infty}\cdots\sum_{n_p=1}^{\infty}\delta_{n_1+\cdots+n_p,N}\delh^{(n_1)}\Gamma \delh^{(n_2)}\cdots\Gamma \delh^{(n_{p-1})}\Gamma H^{(n_p)}\ket{0}
		\end{aligned}
	\end{equation}
	respectively, where $\delta_{i,j}$ is the Kronecker delta enforcing the constraint that the integers $n_m$ sum to $N$. The sums over all $n_m$, combined with the delta, then create all orderings of each partition of $N$. To apply this in the familiar case of just one perturbation one would take $\delh^{(N)}\mapsto-E^{(N)}Q$ for $N\geq2$.
	
	The two equations above are derived from one matrix equation for the following quantity
	\begin{equation*}
		M^{(N)} = \sum_{p=1}^{N}\sum_{n_1=1}^{\infty}\cdots\sum_{n_p=1}^{\infty}\delta_{n_1+\cdots+n_p,N}\delh^{(n_1)}\Gamma \delh^{(n_2)}\cdots\Gamma \delh^{(n_{p-1})}\Gamma \delh^{(n_p)},
	\end{equation*}
	where $E^{(N)}= \mel*{0}{M^{(N)}}{0}$ and $\ket*{\psi^{(N)}}=\Gamma M^{(N)}\ket{0}$ relate this quantity to perturbation theory, and $M^{(N)}$ is the subject of the following section.
	Finite examples of the result are provided at various stages in this paper, for example we provide $M^{(N)}$ to fourth order in equation~\eqref{eq:m-0to4}. We provide the single line of code to generate $M^{(N)}$ in the Appendix, as well as the result to eighth order for infinite perturbations; and to tenth order for one perturbation.
	Since $\delh^{(n)}\ket{0}=H^{(n)}\ket{0}$ and $\bra{0}\delh^{(n)}=\bra{0}H^{(n)}$ we can replace $H^{(n_1)}$ by $\delh^{(n_1)}$ and $H^{(n_p)}$ by $\delh^{(n_p)}$ on either end of the formulae~\eqref{eq:res} if we want to, but this includes the sometimes unknown $-E^{(n_1)}$ and $-E^{(n_p)}$ into the description. This is the resolution to the conflict that $M^{(N)}$ depends on $E^{(N)}$. Since both $\mel*{0}{M^{(N)}}{0}$ and $\Gamma M^{(N)}\ket{0}$ do not depend on $E^{(N)}$, the next order can be calculated recursively without self-consistency. Each term in~\eqref{eq:res} can then be written as a product of $p$ $\delh^{(n)}$'s and the connection with the now-known $M^{(N)}$ can then be established.
	
	\subsection{Relation to other work}
	Our approach builds on a long line of developments in perturbation theory. To our knowledge, extending perturbation theory to an infinite number of perturbations is completely new. Integer partitions have been discussed in different contexts before that appear visually similar but are mathematically and conceptually very different. Quantities analogous to our $\delh^{(1)}=H^{(1)}-E^{(1)}\pbar$ have been introduced previously, but we found no analogues of $\delh^{(n)}$ for $n>1$.
	
	Kato’s rigorous treatment \cite{kato1949} established a formal connection between perturbation theory and integer partitions, though his focus was on convergence analysis and degenerate levels, so we do not require the same overheads. Kato's integer partitions control completely different objects and the hierarchical aspect to perturbation theory, in terms of the lower-order corrections, was not introduced. Messiah later presented a simplified restatement of Kato's work in his textbook \cite{messiah2014}, and a reader should start there to appreciate this approach. 
	Brueckner \cite{brueckner1955} identified a quantity analogous to our $\delh^{(1)}$, which organises the hierarchy of corrections, but only for the first-order correction. L\"owdin’s matrix formulation \cite{lowdin1962} for BWPT is analogous to ours and described that integer partitions are relevant.
	He then expands the quantity $V'=V-(E^{(1)}+E^{(2)}+\cdots)$ which is his closest analogue to our $\delh^{(n)}$, and regroups terms to find the energy corrections at each order to find RSPT. He does not provide the formula in terms of the hierarchy of lower-order energies at an arbitrary order. We find the energy corrections directly in terms of previously calculated terms.
	Soliverez \cite{soliverez1969} introduced a shorthand and provided formulae for the contributions up to fifth order, and later Bracci and Picasso (BP) \cite{bracci2012} developed a similar notation. BP highlighted the relevance of integer partitions as well as the hierarchical nature of the theory, and provided rules to generate corrections at arbitrary order, but these constructions remain more cumbersome than our direct sum-over-partitions formulation.
	Each of these contributions serves a complementary purpose: Kato provides insight into convergence and the formal structure of perturbation theory, L\"owdin offers a matrix-based perspective, Messiah delivers a clear pedagogical treatment, and Soliverez and BP facilitate explicit high-order calculations.
	We also provide the matrix formalism and explicit expressions for eigenvalue corrections. Indeed, one can find up to fourth order in the main text, and to a higher order in the Appendix, which can be used directly if one simply wishes to apply the theory.
	
	\subsubsection*{Outline}
	The structure of this paper is as follows.
	In Section~\ref{sec:setup} we introduce the relevant mathematical structures and discuss how integer partitions are generated.
	In Section~\ref{sec:inf} we show how the formulae of the previous section relate to the eigenvalue and eigenvector corrections provided by perturbation theory.
	In Section~\ref{sec:one} we take the limit of one perturbing Hamiltonian and reproduce the familiar Rayleigh-Schr\"odinger perturbation theory.

	\section{Integer partitions and generating functions}\label{sec:setup}
	The mathematical structure of perturbation theory is controlled by the generating functions,
	\begin{align*}
		M(z) \equiv \sum_{n=1}^{\infty}M^{(n)}z^{n},\quad\delh(z)\equiv\sum_{n=1}^{\infty}\delh^{(n)}z^{n}.
	\end{align*}
	Note that $M^{(n)}$ is a matrix and we find that eigenvalue and eigenvector corrections of perturbation theory are provided by its elements. 
	Indeed, later we show that $\mel*{0}{M^{(n)}}{0}$ and $\Gamma M^{(n)}\ket{0}$ provide the eigenvalue and eigenvector corrections respectively.
	The other elements of $M^{(n)}$ do not correspond to either the eigenvalue or eigenvector correction, so they are not relevant here. 
	The central relationship is
	\begin{align}
		M^{(n)} &\equiv \delh^{(n)} + \sum_{n'=1}^{n-1}\delh^{(n')}\Gamma M^{(n-n')}\label{eq:m-recur-finite}\\
		M(z) &\equiv \delh(z)+\delh(z)\Gamma  M(z)\label{eq:m-recur-infinite}
	\end{align}
	where the first line defines the quantity in the recurrence relationship and the second line is found by converting the first line into generating functions (multiplying by $z^n$, summing $\sum_{n=1}^{\infty}$ and recognising the convolution structure that can be turned into a product of generating functions).
	A simple rearrangement of equation~\eqref{eq:m-recur-infinite} provides an explicit formula for $M(z)$,
	\begin{equation*}
		\begin{aligned}
			M(z) &=(1-\delh(z)\Gamma )^{-1}\delh(z)=\sum_{n=0}^{\infty}[\delh(z) \Gamma]^n\delh(z)
		\end{aligned}
	\end{equation*}
	which allows us to write the explicit formula for a finite order,
	\begin{equation}\label{eq:m-explicit}
		M^{(N)} = \sum_{p=1}^{N}\sum_{n_1=1}^{\infty}\cdots\sum_{n_p=1}^{\infty}\delta_{n_1+\cdots+n_p,N}\delh^{(n_1)}\Gamma \delh^{(n_2)}\cdots\Gamma \delh^{(n_{p-1})}\Gamma \delh^{(n_p)},
	\end{equation}
	that we connect to perturbation theory in the next section.
	
	The explicit relation~\eqref{eq:m-explicit} provides
	\begin{equation}\label{eq:m-0to4}
		\begin{aligned}
			M^{(1)} &= \delh^{(1)}\\
			M^{(2)} &= \delh^{(2)}+\delh^{(1)}\Gamma \delh^{(1)}\\
			M^{(3)} &= \delh^{(3)} + \delh^{(1)}  \Gamma  \delh^{(2)} + \delh^{(2)}  \Gamma  \delh^{(1)} + \delh^{(1)}  \Gamma  \delh^{(1)}  \Gamma  \delh^{(1)}\\		
			M^{(4)} &= \delh^{(4)} + \delh^{(1)}  \Gamma  \delh^{(3)}+ \delh^{(2)}  \Gamma  \delh^{(2)} + \delh^{(3)}  \Gamma  \delh^{(1)}\\
			&\quad + 
			\delh^{(1)}  \Gamma  \delh^{(1)}  \Gamma  \delh^{(2)}+ \delh^{(1)}  \Gamma  \delh^{(2)}
			\Gamma  \delh^{(1)}+ \delh^{(2)}  \Gamma  \delh^{(1)}  \Gamma  \delh^{(1)}\\
			&\quad + 
			\delh^{(1)}  \Gamma  \delh^{(1)}  \Gamma  \delh^{(1)}  \Gamma  \delh^{(1)}
		\end{aligned}
	\end{equation}
	to fourth order, and we see the $2^{n-1}$ terms in $M^{(n)}$ corresponding to all ordered partitions of $n$. This could also have been achieved by recursively substituting lower-order expressions into the recursive definition of $M^{(n)}$ above.
	When calculating $E^{(N)}$, all terms with exterior $E^{(n)}\pbar$ vanish as the starting state is projected out, so that external $\delh^{(n)}$'s map to $H^{(n)}$.
	Our matrix $M$ encodes information required at higher orders in the expansion and thus appears initially more complicated than the results of BP.
	
	To make the sum over integer partitions even clearer we introduce the shorthand
	\begin{equation}\label{eq:symbol}
		\begin{aligned}
			\delh^{(n_1)}\Gamma \delh^{(n_2)}\cdots\Gamma\delh^{(n_p)}&\mapsto (n_1n_2\dots n_p)\\
			\delh^{(n_1)}&\mapsto (n_1)
		\end{aligned}
	\end{equation}
	which has reflection symmetry corresponding to Hermiticity. For Hermitian $H^{(n)}$, the expectation value of a symbol in the initial state equals that of its reversed version, providing a factor of two. This symbol is unambiguous for integers less than 11, the integer 10 is still unambiguous in this symbol because there is no $\delh^{(0)}$. With this equation~\eqref{eq:m-explicit} becomes
	\begin{align*}
		M^{(N)} &= \sum_{p=1}^{N}\sum_{n_1=1}^{\infty}\cdots\sum_{n_p=1}^{\infty}\delta_{n_1+\cdots+n_p,N}(n_1\,n_2\dots n_p)
	\end{align*}
	and the above formulae~\eqref{eq:m-0to4} can be written as
	\begin{align*}
		M^{(1)} &= (1) \\
		M^{(2)} &= (2)+(11) \\
		M^{(3)} &= (3)+(12)+(21)+(111) \\
		M^{(4)} &= (4)+(13)+(22)+(31)\\
		&\quad+(112)+(121)+(211)+(1111).
	\end{align*}
	Code to generate $M^{(N)}$ can be found in the Appendix as well as the result to eighth order for infinite perturbations; and to tenth order for one perturbation.
	
	Equation~\eqref{eq:m-explicit} is the main result of this section, that is, the formula for $M^{(N)}$ as a sum over all ordered partitions of $N$. This provides all possible products of $\delh^{(n)}$, $n\leq N$, separated by $\Gamma$ factors, where the superscripts add up to $N$.
	Note that one requires $E^{(n)}$ to construct $\delh^{(n)}$ for subsequent calculations but one can find $E^{(n)}$ from only $H^{(n)}$.
	We view equation~\eqref{eq:m-explicit} as the formula that is useful if one is going to generate perturbation theory for one or more perturbations, although both the recurrence relation~\eqref{eq:m-recur-infinite} and the explicit formula~\eqref{eq:m-explicit} provide equivalent results. 
	
	\section{Infinite perturbations}\label{sec:inf}
	In this section we connect equation~\eqref{eq:m-explicit} to perturbation theory for an infinite set of perturbations. Although we have infinite perturbations, we isolate the eigenvalue and eigenvector corrections using the same steps as for just one perturbation, as found in any textbook on quantum mechanics.
	We solve equation~\eqref{eq:def} at each order in $z$ to find the corrections to the eigenvalue and eigenvector. At zeroth order in $z$ we find
	\begin{align*}
		H_0\ket*{0}=\epsilon_0\ket*{0},
	\end{align*}
	and at $n$'th order in $z$, $n>0$, we find
	\begin{align*}
		H^{(n)}\ket{0}+H_0\ket*{\psi^{(n)}}+\sum_{n'=1}^{n-1}H^{(n')}\ket*{\psi^{(n-n')}}
		=E^{(n)}\ket{0}+\epsilon_0\ket*{\psi^{(n)}}+\sum_{n'=1}^{n-1}E^{(n')}\ket*{\psi^{(n-n')}}.
	\end{align*}
	To find the corrections to the eigenvalues and eigenvectors we project with $\bra{0}$ and $\bra*{n_1}$ respectively to find,
	\begin{align*}
		\expval*{H^{(n)}}{0}+\sum_{n'=1}^{n-1}\mel*{0}{H^{(n')}}{\psi^{(n-n')}}&=E^{(n)},\\
		\hspace{-0.1em}\mel*{n_1}{H^{(n)}}{0}+\epsilon_{n_1}\braket*{n_1}{\psi^{(n)}}+\sum_{n'=1}^{n-1}\mel*{n_1}{H^{(n')}}{\psi^{(n-n')}}&=\epsilon_{0}\braket*{n_1}{\psi^{(n)}}\\
		&\quad+\sum_{n'=1}^{n-1}\mel*{n_1}{E^{(n')}}{\psi^{(n-n')}},
	\end{align*}
	where we used that $\braket*{0}{\psi^{(n)}}=0$ to simplify the first line, and $\bra{n_1}H_0=\bra{n_1}\epsilon_{n_1}$ to simplify the second line.
	
	A simple rearrangement isolates the relevant quantities
	\begin{align}
		E^{(n)} &= \expval{H^{(n)}}{0}+ \sum_{n'=1}^{n-1}\mel*{0}{H^{(n')}}{\psi^{(n'-n)}}\label{eq:en}
	\end{align}
	and
	\begin{align}
		(\epsilon_0&-\epsilon_{n_1})\braket*{n_1}{\psi^{(n)}}= \mel*{n_1}{H^{(n)}}{0}+ \sum_{n'=1}^{n-1}\mel*{n_1}{\delh^{(n')}}{\psi^{(n'-n)}},\label{eq:psin}
	\end{align}
	where we recognised the previous quantity $\delh^{(n)}$, to further simplify the equations.
	
	The recurrence relationships for the eigenvalue and eigenvector correction can be recognised as elements of the matrix recurrence relation~\eqref{eq:m-recur-finite}. If we make the connection for the seed terms
	\begin{align*}
		E^{(1)}&=\mel*{0}{M^{(1)}}{0},\\
		(\epsilon_0-\epsilon_{n_1})\braket*{n_1}{\psi^{(1)}}&=\mel*{n_1}{M^{(1)}}{0}
	\end{align*}
	then the uniqueness of the recurrence relationship allows us to identify the relationship in general
	\begin{align*}
		E^{(n)}&=\mel*{0}{M^{(n)}}{0},\\ 
		(\epsilon_0-\epsilon_{n_1})\braket*{n_1}{\psi^{(n)}}&=\mel*{n_1}{M^{(n)}}{0}.
	\end{align*}
	By substituting the second line into $\ket*{\psi^{(n)}}=\pbar\ket*{\psi^{(n)}}$ we can write the eigenvector corrections in terms of $M^{(n)}$,
	\begin{align*}
		\ket*{\psi^{(n)}}=\pbar\ket*{\psi^{(n)}}=\sum_{n_1\neq 0}\ket{n_1}\braket*{n_1}{\psi^{(n)}}=\sum_{n_1\neq 0}\frac{\dyad{n_1}}{\epsilon_{0}-\epsilon_{n_1}}M^{(n)}\ket{0}=\Gamma M^{(n)}\ket{0}.
	\end{align*}
	The final step is to substitute this relation into the right-hand sides of equations~\eqref{eq:en} and~\eqref{eq:psin} which provides
	\begin{align*}
		E^{(n)}&=\mel*{0}{\delh^{(n)}}{0}+\sum_{n'=1}^{n-1}\bra{0}\delh^{(n')}\Gamma M^{(n-n')}\ket{0}=	\mel*{0}{M^{(n)}}{0},\\	(\epsilon_0-\epsilon_{n_1})\braket*{n_1}{\psi^{(n)}}&=\mel*{n_1}{\delh^{(n)}}{0}+\sum_{n'=1}^{n-1}\bra{n_1}\delh^{(n')}\Gamma M^{(n-n')}\ket{0}=\mel*{n_1}{M^{(n)}}{0}
	\end{align*}
	respectively as two elements of the matrix equation~\eqref{eq:m-recur-finite}, and we used $\delh^{(n)}\ket{0}=H^{(n)}\ket{0}$ and $\bra{0}\delh^{(n)}=\bra{0}H^{(n)}$ to write the right-hand sides in terms of this one quantity.
	The solution to the matrix equation is~\eqref{eq:m-explicit}, and is provided to fourth order in~\eqref{eq:m-0to4}; in the Appendix we provide up to eighth order for infinite perturbations and up to tenth order for one perturbation.
	
	In practice one does not require the full matrix $M^{(N)}$ but only the vector $M(z)\ket{0}=H(z)\ket{0}+\delh(z)\Gamma M(z)\ket{0}$. This would be of practical benefit, for example,  when studying many-body problems in physics where the Hamiltonian matrix can often be $2^N\times2^N$ for $N$ particles/spins. Dealing with the vector massively reduces the required storage space; then one requires to apply a sparse matrix to this vector.
	
	The correction to the eigenvalue at order $N$ depends only on the corrections at order $N-2$ and below, not on the order $N-1$ term.
	Terms of order $N-1$ only appear on either end of a product of $\delh^{(n)}$ (for example in~\eqref{eq:m-0to4}) and projecting into the starting state on either side annihilates the order $N-1$ eigenvalue corrections to order $N$. The only remaining lower-order contributions are for $-E^{(N-p)}\pbar$ for $p\geq 2$ that appear internally, with at least one $\delh^{(n)}$ on either end. However, one does require the result at order $N-1$ to determine the correction to the eigenvector, since one projects onto the vector with $\ket{n_1}$ and this does not annihilate the energy in $\delh^{(N-1)}$.
	
	\section{One perturbation}\label{sec:one}
	In this section we take the limit that is most familiar in physics, the case of a single perturbing Hamiltonian, and demonstrate that it provides standard perturbation theory. 

	Crucially we use $\delh^{(1)}=H^{(1)}-E^{(1)}\pbar$ instead of $H^{(1)}$ and $-E^{(1)}$ separately which greatly reduces the number of terms in the final formulae and allows us to use the simple sum over all ordered integer partitions to generate the theory. We can recognise this as the central quantity as it appear naturally in the infinite order analogue. In this limit we also have $\delh^{(N)}\mapsto -E^{(N)}\pbar$ for $N\geq2$.
	The analogues of~\eqref{eq:m-0to4} are remarkably simple in this case since any $E^{(n)}\pbar$ on the edges are annihilated by projection and there are no higher-order perturbations. This provides
	\begin{align*}
		PM^{(1)}P=E^{(1)}P &= P(1)P \\
		PM^{(2)}P=E^{(2)}P &= P(11)P \\
		PM^{(3)}P=E^{(3)}P &= P(111)P \\
		PM^{(4)}P=E^{(4)}P &= P(1111)P+P(121)P
	\end{align*}
	to fourth order, in the shorthand notation introduced in~\eqref{eq:symbol}.
	We next show how to go from our shorthand symbol~\eqref{eq:symbol} to the familiar statement of perturbation theory to fourth order.
	Writing this in the original notation we have
	\begin{equation*}
		\begin{aligned}
			E^{(1)} &= \mel*{0}{H^{(1)}}{0}\\
			E^{(2)} &= \mel*{0}{H^{(1)}\Gamma H^{(1)}}{0}\\
			E^{(3)} &= \mel*{0}{H^{(1)}  \Gamma  \delh^{(1)}  \Gamma  H^{(1)}}{0}\\		
			E^{(4)} &= \bra{0}[H^{(1)}  \Gamma  \delh^{(1)}  \Gamma  \delh^{(1)}  \Gamma  H^{(1)}-H^{(1)}  \Gamma  E^{(2)}\Gamma  H^{(1)} ]\ket{0}
		\end{aligned}
	\end{equation*}
	where we used $\pbar \Gamma=\Gamma$ in the final term, and this can be easily written in the familiar form by expanding out the $\delh^{(1)}$'s and using $\Gamma$'s spectral representation,
	\begin{equation*}
		\begin{aligned}
			\expval*{M^{(1)}}{0} &= E^{(1)} = H^{(1)}_{00}\\
			\expval*{M^{(2)}}{0} &= E^{(2)} = \sum_{n_1\neq 0}\frac{H^{(1)}_{0n_1}H^{(1)}_{n_10}}{\epsilon_0-\epsilon_{n_1}}\\
			\expval*{M^{(3)}}{0} &= E^{(3)} =\sum_{\substack{n_1\neq0\\ n_2\neq0}}\frac{H^{(1)}_{0n_1}H^{(1)}_{n_1n_2}H^{(1)}_{n_20}}{(\epsilon_0-\epsilon_{n_1})(\epsilon_0-\epsilon_{n_2})}-E^{(1)}\sum_{n_1\neq 0}\frac{H^{(1)}_{0n_1}H^{(1)}_{n_10}}{(\epsilon_0-\epsilon_{n_1})^2}\\
			\expval*{M^{(4)}}{0} &= E^{(4)} =\sum_{\substack{n_1\neq0\\ n_2\neq0\\ n_3\neq 0}}\frac{H^{(1)}_{0n_1}H^{(1)}_{n_1n_2}H^{(1)}_{n_2n_3}H^{(1)}_{n_30}}{(\epsilon_0-\epsilon_{n_1})(\epsilon_0-\epsilon_{n_2})(\epsilon_0-\epsilon_{n_3})}-E^{(1)}\sum_{\substack{n_1\neq0\\ n_2\neq0}}\frac{H^{(1)}_{0n_1}H^{(1)}_{n_1n_2}H^{(1)}_{n_10}}{(\epsilon_0-\epsilon_{n_1})(\epsilon_0-\epsilon_{n_2})^2}\\
			&\hspace{-1em}-E^{(1)}\sum_{\substack{n_1\neq0\\ n_2\neq0}}\frac{H^{(1)}_{0n_1}H^{(1)}_{n_1n_2}H^{(1)}_{n_10}}{(\epsilon_0-\epsilon_{n_1})^2(\epsilon_0-\epsilon_{n_2})}+[E^{(1)}]^2\sum_{n_1\neq 0}\frac{H^{(1)}_{0n_1}H^{(1)}_{n_10}}{(\epsilon_{0}-\epsilon_{n_1})^3}-E^{(2)}\sum_{n_1\neq 0}\frac{H^{(1)}_{0n_1}H^{(1)}_{n_10}}{(\epsilon_0-\epsilon_{n_1})^2}
		\end{aligned}
	\end{equation*}
	where we used $H^{(1)}_{n_1n_2}\equiv\mel*{n_1}{H^{(1)}}{n_2}$ as shorthand for matrix elements.
	
	\section{Conclusion}
	
	We have presented explicit, systematic formulae in equation~\eqref{eq:res} that provide straightforward access to higher orders of perturbation theory for an arbitrary number of perturbing Hamiltonians, with the single-perturbation case recovered as a limit.  
	At order $N$, one simply partitions the integer $N$ into its $2^{N-1}$ ordered partitions, which determine the combination of $\delh^{(n)}$ in each term.
	This formulation simplifies the otherwise cumbersome derivations of perturbation theory and provides a straightforward formula for high-order calculations.
	
	The subject of future work is when these perturbing Hamiltonians are specified by the power series representation of the Baker-Campbell-Hausdorff formula \cite{moodie2021,jones2025a}, and their internal relationships provide dramatic simplifications \cite{jones2025b}.
	One could then use the resulting perturbation theory to investigate phase transitions in classical statistical mechanics, where the transfer matrix method provides a natural connection to the Baker-Campbell-Hausdorff formula. 
	The same formalism may also prove useful in other areas of physics where the Baker-Campbell-Hausdorff formula arises \cite{scharf1988,dalessio2013,vajna2018}.
	
	\subsubsection*{Acknowledgements}
	We thank Richard Mason, David Reid and Tom Sheppard for comments on the manuscript, and J. M. F. Gunn for useful discussions.
	J.J. is funded by an EPSRC studentship and the University of Birmingham.
	
	\appendix
	\section*{Appendix. Code to generate the main result at a finite order and higher order examples}
	In this appendix we provide some code that can be evaluated in Mathematica. The first two lines of code each generate the required terms for $N$'th order perturbation theory, the final line of code provides the limiting case of one perturbation.	
	Then we provide the explicit results for fifth to eighth order for infinite perturbations and to tenth order for one perturbation. One could then compare the results to seventh order for one perturbation to that of BP \cite{bracci2012}.
	
	The recurrence relationship~\eqref{eq:m-recur-finite} can be encoded in Mathematica as
	\begin{Verbatim}
		m[n_] := dH[n] + TensorExpand[Sum[dH[p] . g . m[n - p], {p, 1, n - 1}]]
	\end{Verbatim}
	where \texttt{dH[n]} and \texttt{g} represent $\delh^{(n)}$ and $\Gamma$ respectively, and \verb|m[n_]| can be called to provide $M^{(n)}$. The following code generates the result in terms of the shorthand symbol,
	\begin{Verbatim}
		symbols[n_] := Flatten[Map[Permutations, IntegerPartitions[n]], 1]
	\end{Verbatim}
	generates the full list of shorthand symbols at order $n$, whose linear combination provides $M^{(n)}$ in shorthand notation.
	For one perturbing Hamiltonian, the following line of code generates $PM^{(n)}P$,
	\begin{Verbatim}
		onePert[n_] := Join[{1}, #, {1}] & /@ symbols[n - 2]
	\end{Verbatim}
	which requires the \verb|symbols| function defined above.
	
	Now we provide the explicit results to a higher order than in the main text.
	We start with the matrix at fifth to eighth order for infinite perturbations,
	\begin{align*}
		M^{(5)} &= (5) + (14) + (23) + (32) + (41) + (113) + (122) + (131) + (212) + (221) + (311)+ (1112)\\
		&\quad +(1121) +(1211) + (2111) + (11111)\\
		M^{(6)} &= (6) + (15) + (24) + (33) + (42) + (51) + (114) +(123) + (132) + (141) + (213) + (222)\\
		&\quad + (231) + (312) + (321) + (411) + (1113) +(1122) + (1131) + (1212)  + (1221) + (1311)\\
		&\quad + (2112) + (2121)+ (2211) + (3111) + (11112)  + (11121) + (11211) + (12111) + (21111)\\
		&\quad + (111111)\\
		M^{(7)} &= (7) + (16) + (25) + (34) + (43) + (52) + (61) + (115) + (124) + (133) + (142)+ (151)\\
		&\quad + (214) + (223) + (232) + (241) + (313) + (322) + (331) + (412) + (421) + (511) + (1114)\\
		&\quad +(1123)+ (1132) + (1141) + (1213) + (1222) + (1231)   + (1312) + (1321) + (1411)\\
		&\quad + (2113) + (2122)+ (2131) + (2212) + (2221)   + (2311) + (3112) + (3121) + (3211)\\
		&\quad + (4111) + (11113) + (11122) + (11131) + (11212) + (11221) + (11311) + (12112)\\
		&\quad + (12121) + (12211) + (13111) + (21112)+ (21121) + (21211) + (22111) + (31111)\\
		&\quad + (111112)  + (111121) + (111211) + (112111) + (121111)+ (211111) + (1111111)\\
		M^{(8)} &=(8) + (44) + (161) + (242) + (323) + (1331) + (2222) + (3113) + (11411) + (12221)\\
		&\quad + (21212)+ (112211) + (121121) + (211112) + (1112111) + (11111111) + \bigg\{(17) + (26)\\
		&\quad + (35) + (116) + (125) + (134) + (143) + (152) + (215) + (224) + (233) + (314)\\
		&\quad + (1115) + (1124) + (1133) + (1142) + (1151) + (1214) + (1223) + (1232) + (1241)\\
		&\quad + (1313)  + (1322) + (1412) + (2114) + (2123) + (2132) + (2213) + (11114) + (11123)\\
		&\quad + (11132) + (11141) + (11213) + (11222) + (11231) + (11312) + (11321) + (12113)\\
		&\quad + (12122) + (12131) + (12212) + (13112) + (21113) + (21122) + (111113) + (111122)\\
		&\quad + (111131) + (111212) + (111221) + (111311) + (112112) + (112121) + (121112)\\
		&\quad + (1111112) + (1111121) + (1111211) + \text{ reversed symbols}\bigg\}
	\end{align*}
	where we employed the reversal symmetry of the symbol to write eighth order, and the result at each order is nothing more than the sum over all ordered partitions of $N$ for $M^{(N)}$ as stated in the main text.
	
	For just one perturbation there are some simplifications. In the shorthand notation, terms with $(n \dots)$ or $(\dots n)$ for $n>1$ vanish under projection with $P$, and $\delh^{(n)}\mapsto -E^{(n)}\pbar$ for $n>1$. 
	Note that in this result there are $2^{N-3}$ terms in $PM^{(N)}P$ for $N\geq 3$; which expand out to provide more terms. This is obvious from the fact that we have two constraints; there must be a $(1,\dots)$ on the left side and $(\dots,1)$ on the right, and the relevant combination of middle integers in the symbol are identical to those in $M^{(N-2)}$.
	We project onto the starting state to eliminate many terms that do not contribute at a given order, providing
	\begin{align*}
		PM^{(5)}P=E^{(5)}P &=P[(131)+(1121) + (1211) + (11111)]P\\
		PM^{(6)}P=E^{(6)}P &= 	P[(141) + (1131) + (1221) + (1311)\\
		&\quad + (11121) + (11211) + (12111) + (111111)]P\\
		PM^{(7)}P=E^{(7)}P &= P[(151) + (1141) + (1231)+(1321) + (1411)+ (11131)\\
		&\quad + (11221) + (11311)+ (12121) + (12211)+(13111)\\
		&\quad + (111121) + (111211) + (112111) + (121111) + (1111111)]P.
	\end{align*}
	as the analogue of BP's results to seventh order, note that BP use $D$ as the denominator matrix which is just the negative of our $\Gamma$, so that their denominators are positive numbers.
	Eighth order is,
	\begin{align*}
		PM^{(8)}P=E^{(8)}P&= P[(161) +(1151) + (1241) + (1331) + (1421) + (1511) + (11141) + (11231)\\
		&\quad + (11321)+ (11411) + (12131) +(12221) + (12311) + (13121) + (13211)\\
		&\quad + (14111) +(111131)+ (111221) + (111311) + (112121) + (112211)\\
		&\quad + (113111) + (121121) + (121211)+ (122111) + (131111) + (1111121)\\
		&\quad + (1111211) +(1112111) + (1121111) +(1211111)+ (11111111)]P.
	\end{align*}
	For ninth and tenth order we employ the reversal symmetry in our symbol
	\begin{align*}
		PM^{(9)}P=E^{(9)}P &= P\bigg[(171) + (11511) + (12321) + (13131) + (1113111) + (1121211)\\
		&\quad + (1211121)+ (111111111)+ \bigg\{ (1161) + (1251) + (1341) + (11151)\\
		&\quad + (11241) + (11331)+ (11421) + (12141) + (12231) + (111141) + (111231)\\
		&\quad + (111321) + (111411)+ (112131) + (112221) + (112311) + (113121)\\
		&\quad + (121131)  + (121221) + (1111131) + (1111221) + (1111311) + (1112121)\\
		&\quad + (1112211) + (1121121) + (11111121)+ (11111211) + (11112111)\\
		&\quad + \text{ reversed symbols}\bigg\}	\bigg]P
	\end{align*}
	\begin{align*}
		PM^{(10)}P=E^{(10)}P &= P\bigg[(181) + (1441) + (11611) + (12421) + (13231) + (113311) + (122221)\\
		&\quad + (131131) + (1114111) + (1122211) + (1212121) + (11122111)\\
		&\quad + (11211211) + (12111121) + (111121111) + (1111111111) + \bigg\{(1171)\\
		&\quad + (1261) + (1351) + (11161) + (11251) + (11341) + (11431) + (11521)\\
		&\quad + (12151) + (12241) + (12331) + (13141) + (111151) + (111241) + (111331)\\
		&\quad + (111421) + (111511) + (112141) + (112231) + (112321) + (112411)\\
		&\quad + (113131) + (113221) + (114121) + (121141) + (121231) + (121321)\\
		&\quad + (122131) + (1111141) + (1111231) + (1111321) + (1111411) + (1112131)\\
		&\quad + (1112221) + (1112311) + (1113121) + (1113211) + (1121131) + (1121221)\\
		&\quad + (1121311) + (1122121) + (1131121) + (1211131) + (1211221) + (11111131)\\
		&\quad + (11111221) + (11111311) + (11112121) + (11112211) + (11113111)\\
		&\quad + (11121121) + (11121211) + (11211121) + (111111121) + (111111211)\\
		&\quad + (111112111) + \text{ reversed symbols}\bigg\}\bigg]P.
	\end{align*}
	
	\bibliographystyle{unsrt}
	\bibliography{refs.bib}
	
\end{document}